\documentclass[british,prerpint, preprintnumbers, final]{revtex4-2}
\PassOptionsToPackage{obeyFinal}{todonotes}
\usepackage[T1]{fontenc}
\usepackage[latin9]{inputenc}
\usepackage{color}
\usepackage{todonotes}
\usepackage{amsmath}
\usepackage{graphicx}
\usepackage{esint}

\makeatletter
\date{}

\usepackage{xcolor}
\newcommand{\chnga}[1]{{#1}}

\makeatother

\usepackage{babel}
\begin{document}
\global\long\def\mpl{m_{\mathrm{Pl}}}%

\global\long\def\inf{\mathrm{I}}%

\global\long\def\eq{\mathrm{eq}}%

\global\long\def\r{\mathrm{r}}%

\global\long\def\dm{\mathrm{DM}}%

\global\long\def\reh{\mathrm{reh}}%

\global\long\def\fr{\mathrm{fi}}%
 
\global\long\def\fri{\fr}%

\global\long\def\td{\mathrm{td}}%

\global\long\def\rd{\mathrm{rd}}%

\global\long\def\gw{\mathrm{GW}}%

\global\long\def\bbn{\mathrm{BBN}}%

\global\long\def\dec{\mathrm{dec}}%

\global\long\def\s{\phi}%

\preprint{IPARCOS-UCM-23-124}
\title{Constraining the General Oscillatory Inflaton Potential with Freeze-in
Dark Matter and Gravitational Waves}
\author{Jose A. R. Cembranos}
\affiliation{Departamento de Física Teórica and IPARCOS, Facultad de Ciencias Físicas,
Universidad Complutense de Madrid, Ciudad Universitaria, 28040 Madrid,
Spain}
\author{Mindaugas Kar\v{c}iauskas}
\affiliation{Departamento de Física Teórica, Facultad de Ciencias Físicas, Universidad
Complutense de Madrid, Ciudad Universitaria, 28040 Madrid, Spain}
\affiliation{Center for Physical Sciences and Technology, Saul\.{e}tekio av. 3,
10257 Vilnius, Lithuania}
\begin{abstract}
The reheating phase after inflation is one of the least observationally
constrained epochs in the evolution of the Universe. The forthcoming
gravitational wave observatories will enable us to constrain at least
some of the non-standard scenarios. For example, models where the
radiation bath is produced by the perturbative inflaton decay that
oscillates around a minimum of the potential of the form $V\propto\phi^{2n}$,
with $n>2$. In such scenarios a part of the inflationary gravitational
wave spectrum becomes blue tilted, making it observable, depending
on the inflation energy scale and the reheating temperature. The degeneracy
between the latter two parameters can be broken if dark matter in
the Universe is produced via the freeze-in mechanism. The combination
of the independent measurement of dark matter mass with gravitational
wave observations makes it possible to constrain the reheating temperature
and the energy scale at the end of inflation, at least within some
parameter ranges.
\end{abstract}
\maketitle

\section{Introduction}

Freeze-out is a popular mechanism to explain the production of Dark
Matter (DM) \citep{Lee:1977ua,Scherrer:1985zt,Srednicki:1988ce,Gondolo:1990dk,Kolb:1990vq}.
According to this mechanism DM is in thermal equilibrium with the
hot plasma of the early Universe. As the Universe expands and cools
down, the DM interaction rate becomes smaller than the expansion rate.
This makes DM particles decouple from the rest of plasma, leaving
the comoving particle number density constant thereafter. The final
DM abundance in this scenario is not very sensitive to the physics
of the early Universe; at least not for models with a standard thermal
history after the freeze-out.

An alternative DM production mechanism is the so called ``freeze-in''
mechanism (see e.g. \citep{Chung:1998rq}). According to this paradigm,
the interactions of DM are so weak that they are never in thermal
equilibrium with the cosmic plasma. However, such interactions are
strong enough that the scattering of the thermal bath particles can
produce DM in sufficient abundance. After the production, DM particles
travel unimpeded. In contrast to the freeze-out scenario, the DM abundance
from the freeze-in scenarios is very sensitive to the details of the
early Universe physics. This offers an opportunity: if Nature did
select this scenario for DM production, by measuring DM properties
will enable us to learn and constrain the physics of the early Universe.

The potential of such studies is, for example, investigated in Refs.~\citep{Feldstein:2013uha,Bhattiprolu:2022sdd},
where it is demonstrated how the measurement of DM particle mass could
give us information about the maximum themperature and the reheating
temperature after inflation. Any new method to constraint the epoch
of reheating by observations is very welcome, as this remains to be
a very poorly understood period due to the lack of direct observational
signatures. The authors of Ref.~\citep{Feldstein:2013uha} assumed
hypercharged minimal DM and a standard reheating scenario. According
to such a scenario the inflaton oscillates around the minimum of the
quadratic potential after inflation and perturbatively decays into
relativistic particles \citep{Lyth:2009zz}. The sensitivity of the
freeze-in mechanism to the early Universe physics also means that
any change in this standard reheating scenario could affect the properties
of DM particles. For example, the effects of the non-standard expansion
history after inflation is studied in Refs.~\citep{DEramo:2017ecx,Bernal:2020bfj,Ghoshal:2022ruy,Barman:2022qgt,Barman:2023ktz}
and the authors of Ref.~\citet{Becker:2023tvd} investigate other
ways how freeze-in DM could be used to constrain the reheating epoch. 

In the past several years the detection of Gravitational Waves (GW)
\citep{LIGOScientific:2016aoc,Abbott:2016nmj,LIGOScientific:2017vwq}
opened a new window to investigate the Universe. More recently, the
detection of the stochastic GW background by pulsar timing array collaborations
NANOGrav, EPTA/InPTA, PPTA and CPTA has reached another big milestone
in the GW astronomy \citep{NANOGrav:2023gor,Xu:2023wog,EPTA:2023fyk,Reardon:2023gzh}.
Although the detected background signal is generated by astrophysical
sources, it demonstrates the potential to study the physics of the
early Universe too. There are numerous processes that could operate
in the early Universe and produce a detectable background of GW \citep{Guzzetti:2016mkm,Caprini:2018mtu}.
One of such processes is the non-standard expansion history after
inflation \citep{Giovannini:1998bp,Giovannini:1999bh,Figueroa:2019paj,Haque:2021dha}.
Due to vacuum fluctuations, inflation leaves the Universe filled with
the primordial GW background that is (quasi) scale invariant \citep{Lyth:2009zz}.
This spectrum is modified as the Universe undergoes the various phases
of its evolution \citep{Saikawa:2018rcs}. In particular, any epoch,
where the expansion rate deviates from the radiation dominated one,
tilts the spectrum in the range of subhorizon modes. In many models
the tilt can be large enough that the inflationary GW signal falls
within the sensitivity limits of future observatories (see for example
\citep{Haque:2021dha,Dimopoulos:2023izi}).

In this work we investigate how the combination of DM searches and
GW observations could shed some light on inflation and the reheating
epochs. In Section~\ref{sec:production} we compute the freeze-in
production of DM particles during reheating. In contrast to the previous
calculations we allow for the inflaton to oscillate around a non-quadratic
potential. This modifies the expansion rate during reheating and the
final density of DM. The non-standard expansion rate also affects
the spectrum of the background GW. The computation of such a spectrum
is summarised in Section~\ref{sec:GW}. In Section~\ref{sec:DMGW}
we find the relation between that spectrum and the DM properties.
We also explore the allowed parameter range and determine regions
which are accessible by future observations.

\section{The Freeze-in DM Production with an Arbitrary Inflaton Potential\label{sec:production}}

\subsection{The Setup\label{subsec:Setup}}

We consider a similar setup as discussed in Ref.~\citep{Chung:1998rq}.
The model consists of three components: an inflaton field, radiation
and the non-relativistic DM. After inflation the inflaton $\phi$
oscillates around the minimum of the potential and initially dominates
the energy budget of the Universe. To simplify the discussion, we
assume that the inflaton decays almost exclusively into light degrees
of freedom (radiation) with the decay rate $\Gamma_{\phi}$. The thermalisation
rate is taken to be fast enough so that we can assume these degrees
of freedom to be in a local thermodynamic equilibrium. We can then
relate the energy density of radiation to the thermodynamic temperature
by
\begin{eqnarray}
\rho_{\r} & = & \frac{\pi^{2}g_{*}\left(T\right)}{30}T^{4}\,,\label{rhoT}
\end{eqnarray}
where $g_{*}\left(T\right)$ is the number of effective relativistic
degrees of freedom at a given temperature. The Universe is reheated
with the temperature $T_{\reh}$ when radiation takes over other contributions
to the energy budget. At that moment the Universe becomes radiation
dominated. It is important to recognise that $T_{\reh}$ is not the
maximum temperature, as we will see later.

The heavy and stable DM particles are non-relativistic throughout
the whole of reheating process. Moreover, they are so weakly coupled
to radiation that DM particles never come into thermal equilibrium.
Nevertheless, the interaction is sufficiently strong that the thermal
bath produces DM particles, which are then freely diluted by the expansion
of the Universe.

The Boltzman equations (in terms of energy densities) for such a setup
can be written as \citep{Chung:1998rq}
\begin{eqnarray}
\dot{\rho}_{\phi}+3H\left(1+w_{\phi}\right)\rho_{\phi} & = & -\Gamma_{\phi}\rho_{\phi}\,,\label{EoMinf}\\
\dot{\rho}_{\r}+4H\rho_{\r} & = & \Gamma_{\phi}\rho_{\phi}\,,\label{EoMr}\\
\dot{\rho}_{\dm}+3H\rho_{\dm} & = & \frac{\left\langle \sigma\left|v\right|\right\rangle }{M_{\dm}}\left(\rho_{\dm}^{\eq}\right)^{2}\,,\label{EoMdm}
\end{eqnarray}
where $\rho_{\phi}$, $\rho_{\r}$ and $\rho_{\dm}$ are the energy
densities of the inflaton, radiation and DM respectively. $M_{\dm}$
is the mass of DM particles and $\left\langle \sigma\left|v\right|\right\rangle $
is the thermal average cross section for annihilation of DM. As DM
is assumed to be far from equilibrium, only the $\left(\rho_{\dm}^{\eq}\right)^{2}$
term dominates the R.H.S. of eq.~(\ref{EoMdm}), which is the non-relativistic
equilibrium energy density, given by
\begin{eqnarray}
\rho_{\dm}^{\eq}\left(T\right) & = & \frac{M_{\dm}^{5/2}T^{3/2}}{\left(2\pi\right)^{3/2}}\cdot\mathrm{e}^{-M_{\dm}/T}\,.
\end{eqnarray}
The inflaton decay rate $\Gamma_{\phi}$ is assumed to be much lower
than the Hubble parameter at the end of inflation, $H_{\inf}$, 
\begin{eqnarray}
\Gamma_{\phi} & \ll & H_{\inf}\,.\label{GHinf}
\end{eqnarray}
This leads to a long period of reheating, before the inflaton decays
into radiation completely. 

Equations (\ref{EoMinf})-(\ref{EoMdm}) differ from analogous equations
in Ref.~\citep{Chung:1998rq} by that we do not assume the inflaton
to be oscillating in a quadratic potential after inflation. We rather
allow for a much flatter form of the potential at the minimum, such
that

\begin{eqnarray}
V\left(\phi\right) & \propto & \phi^{2n}\,,\label{V}
\end{eqnarray}
where $n$ is a natural number. Moreover, since we are interested
in observable stochastic background of GWs, we only consider values
$n>2$. These types of potentials can be encountered, for example,
in $\alpha$-attractor \citep{Ueno:2016dim,DiMarco:2017zek} or monodromy
\citep{McAllister:2014mpa} models of inflation. As it is well known
(see \citep{Turner:1983he,Shtanov:1994ce,Lozanov:2016hid,Cembranos:2015oya}),
an oscillating scalar field in such a potential leads to an effective
equation of state of the form
\begin{eqnarray}
w_{\phi} & = & \frac{n-1}{n+1}\,.\label{wn}
\end{eqnarray}
This value of $w_{\phi}$ is used in eq.~(\ref{EoMinf}) above. $n=2$
corresponds to $w_{\phi}=1/3$; as $n$ increases, $w_{\phi}$ approaches
1. However, $w_{\phi}=1$ describes a non-oscillatory inflaton potential
\citep{Felder:1999pv}, hence we cannot assume the standard perturbative
reheating scenario and apply our calculations to estimate DM abundances.
The $w_{\phi}=1$ case rather serves as an upper limit for the allowed
range of $w_{\phi}$ values. With non-oscillatory potentials one has
to employ other ways to reheat the universe. In summary, the range
of $w_{\phi}$ values that will be used in this work is $1/3<w_{\phi}<1$.

Inflaton oscillations in a non-quadratic potential also induce the
time dependence of the decay rate $\Gamma_{\phi}$ \citep{Shtanov:1994ce,Garcia:2020eof,Ahmed:2021fvt,Barman:2022tzk,Banerjee:2022fiw,Barman:2023ktz},
which we parametrise by
\begin{eqnarray}
\Gamma_{\phi} & = & \Gamma_{\inf}a^{\nu}\,,\label{G}
\end{eqnarray}
where $\Gamma_{\inf}$ is the inflaton decay rate at the end of inflation
and we normalised the scale factor to equal unity at that moment $a_{\inf}=1$.
The scaling $\nu$ depends on the shape of the potential in eq.~(\ref{V})
as well as on the dominant inflaton decay channel. According to \citep{Shtanov:1994ce}
if the inflaton decays primarily into fermions then $\nu=-3w_{\phi}$
and $\nu=3w_{\phi}$ if it decays into scalar particles. One can device
a more complicated behaviour of $\Gamma_{\phi}$ (see e.g. \citep{Co:2020xaf})
but we consider only this generic case in the current work.

\subsection{Evolution of the Inflaton and Radiation During Reheating}

The condition in eq.~(\ref{GHinf}) ensures that before the end of
reheating the R.H.S. of eq.~(\ref{EoMinf}) is negligible. Here,
we define the \emph{moment of reheating} to be the moment when radiation
energy density takes over that of the inflaton, i.e. reheating happens
at $a_{\reh}$ when $\rho_{\r}^{\reh}=\rho_{\phi}^{\reh}$. The interval
between inflation and this point we call the \emph{period} \emph{of
reheating}. 

It is also a sufficiently good approximation to assume that the oscillating
inflaton dominates the energy budget of the Universe before radiation
takes over. In other words, we can approximate $\rho\simeq\rho_{\phi}$
during reheating, where $\rho$ is the total energy density. Solving
eq.~(\ref{EoMinf}) then leads to
\begin{eqnarray}
\rho_{\phi} & \simeq & \rho_{\inf}a^{-3\left(1+w_{\phi}\right)}\,,\label{rhophi}
\end{eqnarray}
where $a$ is the scale factor and $\rho_{\inf}$ is the energy density
at the end of inflation. Plugging this result into the Friedman equation
$3\mpl^{2}H^{2}=\rho$ we can also compute the scaling of the Hubble
parameter
\begin{eqnarray}
H & \simeq & H_{\inf}a^{-\frac{3}{2}\left(1+w_{\phi}\right)}\,.\label{H}
\end{eqnarray}

The evolution of the radiation energy density is governed by eq.~(\ref{EoMr}).
To find the solution of this equation during reheating, we can plug
eqs.~(\ref{rhophi}), (\ref{H}) and (\ref{G}) into eq.~(\ref{EoMr}).
At the lowest order in $\Gamma_{\phi}/H\ll1$ the integral is equal
to 
\begin{eqnarray}
\rho_{\r} & = & \frac{\rho_{\inf}}{4\alpha}\frac{\Gamma_{\inf}}{H_{\inf}}\left(a^{4\alpha}-1\right)a^{-4}\,,\label{rhor}
\end{eqnarray}
where $\alpha$ is defined as
\begin{eqnarray}
\alpha & \equiv & \frac{1}{8}\left(5-3w_{\s}+2\nu\right)\,.\label{alph-def}
\end{eqnarray}
With the upper bound $w_{\phi}<1$, that were specified in the paragraph
bellow eq.~(\ref{wn}), and for $-3w_{\s}\le\nu\le3w_{\s}$, the
range of possible  $\alpha$ values is $-1/2<\alpha<1$. Notice,
that the above solution is not valid if $\alpha=0$ ($2\nu=3w_{\s}-5$).
In that case $\rho_{\r}\propto a^{-4}\ln a$. But since we are mainly
interested in $\nu=\pm3w_{\s}$ values, this case will not be analysed
any further. 

Whichever the sign of $\alpha$, the maximum value of $\rho_{\r}$
is achieved at 
\begin{eqnarray}
a_{\max} & = & \left(1-\alpha\right)^{-\frac{1}{4\alpha}}\:.
\end{eqnarray}
After this point the evolution of radiation does depend on the sign
of $\alpha$. In the case of $\alpha>0$, the first term in eq.~(\ref{rhor})
dominates. We call this case the ``lasting inflaton decay case'',
or the ``\emph{lasting decay}'' briefly. In this case the production
of radiation by the oscillating inflaton field continues to be significant
all the way up until reheating is complete. Newly produced radiation
continuously sources the radiation bath and makes its energy density
decrease as $\rho_{\r}\propto a^{-\frac{3}{2}\left(1+w_{\phi}\right)+\nu}$,
i.e. slower than $a^{-4}$.

In the opposite regime $\alpha<0$, which we call the ``\emph{brief
decay}'' case, the radiation production is very inefficient. After
reaching $\rho_{\r}^{\max}$, the energy density of the radiation
bath cannot keep up with the expansion of the Universe and decays
adiabatically, $\rho_{\r}\propto a^{-4}$.

Whichever the decay regime, the lasting or the brief one, we can compute
the maximum temperature of the Universe. If we assume that the number
of effective relativistic degrees of freedom $g_{*}\left(T\right)$
in eq.~(\ref{rhoT}) is constant during reheating, which we denote
by $g_{*}$, the maximum value of $\rho_{\r}$ in eq.~(\ref{rhor})
can be related to the maximum temperature of radiation, which is 
\begin{eqnarray}
T_{\max}^{4} & = & \rho_{\inf}\frac{15}{2\pi^{2}g_{*}}\frac{\Gamma_{\inf}}{H_{\inf}}\left(1-\alpha\right)^{\frac{1}{\alpha}-1}\,.\label{Tmax}
\end{eqnarray}

\begin{figure}
\begin{centering}
\includegraphics[scale=0.6]{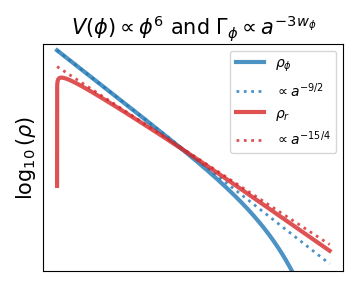}\includegraphics[scale=0.6]{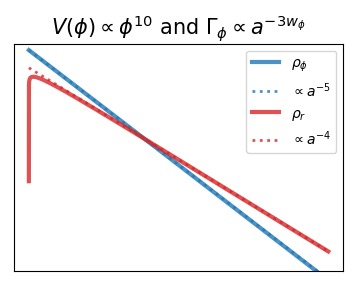}
\par\end{centering}
\begin{centering}
\includegraphics[scale=0.6]{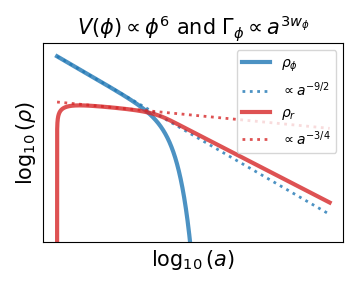}\includegraphics[scale=0.6]{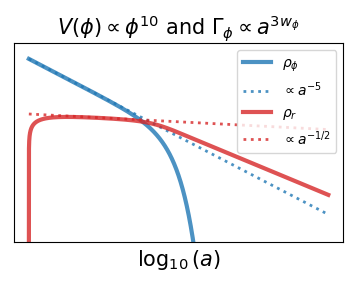}
\par\end{centering}
\caption{\label{fig:rho-reh}The evolution of the inflaton and radiation energy
densities $\rho_{\phi}$ and $\rho_{\protect\r}$ respectively during
reheating. The dotted lines represent approximate scalings derived
in the text.}
\end{figure}

To check if the above described scenario is correct, we also numerically
integrate the exact system of equations (\ref{EoMinf})-(\ref{EoMdm}).
A few examples of those solutions are shown in Figure~\ref{fig:rho-reh}
together with a few lines illustrating analytic solutions in eqs.~(\ref{rhophi})
and (\ref{rhor}). We can clearly see that after the initial burst
of radiation production, the latter passively decays away as $a^{-4}$
in the brief decay regime ($\alpha<0$). This is demonstrated in the
upper row of the figure. Conversely, for $\alpha>0$ (the lower row
of the figure) the radiation production continues to be significant
until the moment of reheating.

As was mentioned above, we define the moment of reheating, to be the
instance when the radiation energy density starts dominating over
the inflaton one. This can happen in both, brief or lasting, decay
scenarios. In the former case, even if the production of radiation
is negligible after $a_{\max}$, the radiation redshifts away slower
than the inflaton (see the upper row of Figure \ref{fig:rho-reh}),
eventually taking it over.

In the lasting decay case several things can happen. First, as in
the previous case, $\rho_{\r}$ might take over $\rho_{\s}$ at $a=a_{\rd}$.
But the evolution of $\rho_{\r}$ was computed assuming $\Gamma_{\phi}/H\ll1$,
which scales as $\Gamma_{\phi}/H\propto a^{\nu+\frac{3}{2}\left(1+w_{\phi}\right)}$.
For the most of parameter values this ratio is growing with time.
It is therefore possible that $\Gamma_{\phi}/H$ becomes larger than
1 at $a=a_{\dec}<a_{\rd}$, before the domination of $\rho_{\r}$
in eq.~(\ref{rhor}). When the $\Gamma_{\phi}<H$ bound is violated,
the inflaton decays away in less than a Hubble time, giving its energy
to radiation. Hence, in the lasting decay scenario we can define the
moment of reheating to be $a_{\reh}=\min\left[a_{\rd},a_{\dec}\right]$.
In practice, however, it is sufficient to use the approximation $a_{\reh}\simeq a_{\rd}$.
To see this, notice that $a_{\rd}/a_{\dec}=\left(4\alpha\right)^{1/\left(4\alpha-1+3w_{\s}\right)}$,
which is never much larger than 1 for the interesting parameter values.
In conclusion, for the rest of the paper the moment of reheating will
be determined using eq.~(\ref{rhor}) and finding the moment $\rho_{\r}\left(a_{\reh}\right)\simeq\rho_{\s}\left(a_{\reh}\right)$.

With this in mind, we find the scale factor at reheating to be
\begin{eqnarray}
a_{\reh} & = & \left(4\left|\alpha\right|\frac{H_{\inf}}{\Gamma_{\inf}}\right)^{\frac{1}{4\vartheta-1+3w_{\phi}}}\,,
\end{eqnarray}
where we introduced the parameter
\begin{eqnarray}
\vartheta & \equiv & \max\left\{ 0,\alpha\right\} \,
\end{eqnarray}
to take into account both cases, $\alpha>0$ and $\alpha<0$. It is
easy to compute the temperature of the Universe at $a_{\reh}$, which
is given by
\begin{eqnarray}
T_{\reh}^{4} & = & \rho_{\inf}\left(\frac{30}{\pi^{2}g_{*}}\right)\left(\frac{\Gamma_{\inf}}{4\left|\alpha\right|H_{\inf}}\right)^{\frac{3\left(1+w_{\phi}\right)}{4\vartheta+3w_{\phi}-1}}\,.\label{Treh}
\end{eqnarray}

\subsection{Freeze-In Production of Dark Matter}

Next, we integrate eq.~(\ref{EoMdm}), which specifies the evolution
of DM. Before the epoch of matter-radiation equality DM is subdominant
to other components of the Universe. This allows us to substitute
$H$ and $T$ by the values computed in the previous section when
we integrate eq.~(\ref{EoMdm}) next. Moreover, for concreteness,
we assume that the DM annihilation is dominated by the s-wave channel,
that is very common for renormalizable interactions. In such a case,
the thermal average cross section is just a constant value. To do
the integration it is first convenient to rewrite eq.~(\ref{EoMdm})
as
\begin{eqnarray}
a^{3}\rho_{\dm}^{\reh} & = & \kappa\intop_{1}^{a}\mathrm{e}^{-g\left(a'\right)}\mathrm{d}a'\,,\label{intg}
\end{eqnarray}
where we used $\rho_{\dm}\left(a_{\inf}=1\right)=0$. The dimension-4
constant $\kappa$ in the above expression depends on the DM cross
section, and is given by
\begin{eqnarray}
\kappa & = & \frac{M_{\dm}^{2}\left\langle \sigma\left|v\right|\right\rangle }{\left(\lambda\pi\right)^{3}}\cdot\frac{M_{\dm}^{5}}{H_{\inf}}\,.
\end{eqnarray}
While the $g\left(a\right)$ function is defined as  
\begin{eqnarray}
g\left(a\right) & \equiv & -\lambda\left(\frac{a^{4\alpha}-1}{4\alpha}\right)^{-\frac{1}{4}}a+\frac{3}{4}\ln\left(\frac{a^{4\alpha}-1}{4\alpha}\right)+\frac{1}{2}\left(1+3w_{\phi}\right)\ln a\,.\label{gy}
\end{eqnarray}
where
\begin{eqnarray}
\lambda & \equiv & \sqrt{2}a_{\max}^{\alpha-1}x_{\max}
\end{eqnarray}
and we introduced a dimensionless parameter
\begin{eqnarray}
x & \equiv & \frac{M_{\dm}}{T}\,,
\end{eqnarray}
with $x_{\max}\equiv x\left(T_{\max}\right)$ and $T_{\max}$ given
in eq.~(\ref{Tmax}).

If the mass of DM particles is sufficiently smaller than the maximal
temperature, $M_{\dm}<T_{\max}$, the integral in eq.~(\ref{intg})
can be evaluated using the Laplace approximation. In that case we
can take the freeze-in scale factor to be $a_{\fr}>1$. The approximate
value of $a_{\fr}$, which depends on the sign of $\alpha$, is 
\begin{eqnarray}
a_{\fr} & \simeq & \left[\frac{6\vartheta+1+3w_{\phi}}{2\sqrt{2}\lambda\left|\alpha\right|^{\frac{1}{4}}\left(1-\theta\right)}\right]^{\frac{1}{1-\vartheta}}\,.
\end{eqnarray}
At $a_{\fr}$ the comoving energy density of DM particles freezes
in and remains constants thereafter. That value is given by
\begin{eqnarray}
a^{3}\rho_{\dm} & = & \frac{16a_{\fri}^{4}}{\pi^{5/2}}\cdot M_{\dm}^{2}\left\langle \sigma\left|v\right|\right\rangle \frac{M_{\dm}^{5}}{H_{\inf}}\left(\frac{1-\vartheta}{6\vartheta+1+3w_{\phi}}\right)^{\frac{7}{2}}\left(\frac{a_{\fri}^{1-\vartheta}}{\mathrm{e}}\right)^{\frac{6\vartheta+1+3w_{\phi}}{2\left(1-\vartheta\right)}}\frac{a_{\fri}^{-3\vartheta}}{1-\vartheta}\,.\label{frin-last}
\end{eqnarray}

Utilising the constancy of $a^{3}\rho_{\dm}$ we can estimate the
present day DM abundance 
\begin{eqnarray}
\frac{\Omega_{\dm}}{\Omega_{\r}} & = & \frac{\rho_{\dm}^{\reh}}{\rho_{\r}^{\reh}}\left[\frac{g_{*}\left(T_{\td}\right)}{g_{*}}\right]^{\frac{1}{3}}\left(\frac{a_{\td}}{a_{\reh}}\right)\,,
\end{eqnarray}
where `$\td$' denotes values evaluated today (remember that $g_{*}\equiv g_{*}\left(T_{\reh}\right)$).
$\Omega_{\dm}$ and $\Omega_{\r}$ denote the current values of DM
and radiation respectively. Plugging eq.~(\ref{rhoT}) into the above
expression we find 
\begin{eqnarray}
\frac{\Omega_{\dm}}{\Omega_{\r}} & = & \frac{30}{\pi^{2}g_{*}}\frac{\rho_{\dm}^{\reh}}{T_{\reh}^{3}T_{\td}}\,.
\end{eqnarray}
Computing $\rho_{\dm}^{\reh}$ from eq.~(\ref{frin-last}) we can
write
\begin{eqnarray}
x_{\reh}\equiv\frac{M_{\dm}}{T_{\reh}} & = & \left[\frac{g_{*}}{80\sqrt{3\pi}}\frac{\Omega_{\r}\mpl}{\Omega_{\dm}T_{\td}}M_{\dm}^{2}\left\langle \sigma\left|v\right|\right\rangle \frac{\mathrm{e}^{-\frac{6\vartheta+1+3w_{\phi}}{2\left(1-\vartheta\right)}}}{1-\vartheta}\cdot\left(\frac{6\vartheta+1+3w_{\phi}}{4\left(1-\vartheta\right)}\right)^{\frac{7\vartheta+2+3w_{\phi}}{2\left(1-\vartheta\right)}}\right]^{\frac{2\left(1-\vartheta\right)}{10\vartheta-1+3w_{\phi}}}\,,\label{xreh}
\end{eqnarray}
where $T_{\reh}$ is given in eq.~(\ref{Treh}). The $\Omega_{\r}\mpl/\Omega_{\dm}T_{\td}$
factor in this expression is fixed by observations. Taking $T_{\td}=2.725\,\text{K}$
\citep{Fixsen:2009ug}, which translates to $\Omega_{\r}h^{2}\simeq2.5\times10^{-5}$,
and $\Omega_{\dm}h^{2}\simeq0.12$ \citep{Planck:2018vyg}, where
$h$ is the dimensionless Hubble parameter today defined as $H_{\td}=100h\text{ km/s/Mpc}$,
we can compute this factor to be 
\begin{eqnarray}
\frac{\Omega_{\dm}}{\Omega_{\r}}\frac{T_{\td}}{\mpl} & \simeq & 4.6\times10^{-28}\,.
\end{eqnarray}
Eq.~(\ref{xreh}) is one of the main equations of this work. It gives
the relation between DM particle mass and the reheating temperature.
For later convenience we can also compute the $x_{\max}$ value, which
is 
\begin{eqnarray}
x_{\max} & = & x_{\reh}\frac{a_{\max}^{1-\alpha}}{\left|\alpha\right|^{\frac{1}{4}}}\left(\frac{\pi^{2}g_{*}}{30}\frac{M_{\dm}^{4}}{x_{\reh}^{4}\rho_{\inf}}\right)^{\frac{1-\vartheta}{3\left(1+w_{\s}\right)}}\,.\label{xmax-xreh}
\end{eqnarray}

\section{Gravitational Waves\label{sec:GW}}

\subsection{The Spectrum}

After inflation is over, the Universe is unavoidably filled with a
background of primordial GWs that have an almost scale invariant spectrum
\citep{Lyth:2009zz}
\begin{eqnarray}
\mathcal{P}_{h} & = & \frac{8}{\mpl^{2}}\left(\frac{H_{k}}{2\pi}\right)^{2}\,,\label{Ph}
\end{eqnarray}
where $H_{k}$ is the Hubble parameter during inflation at the moment
when the $k$ mode exits the horizon. For our purpose it is sufficient
to assume that $H_{k}\simeq\mathrm{const}$ during inflation, which
makes $\mathcal{P}_{h}$ in eq.~(\ref{Ph}) scale invariant. 

After inflation, when those modes reenter the horizon, they behave
as radiation, in particular, they are redshifted according to $\rho_{hk}\left(k<aH\right)\propto a^{-4}$.
On the contrary, the behaviour of the dominant scalar field energy
density depends on the shape of the potential in eq.~(\ref{V}) .
For $n>2$, or equivalently $w_{\phi}>\frac{1}{3}$ in eq.~(\ref{wn}),
the energy density of the oscillating inflaton dilutes faster than
radiation. In those cases the relative contribution of GWs to the
total energy budget of the Universe increases. The earlier the mode
reenters the horizon before the end of reheating, the longer the relative
energy density of that modes grows. This modifies the primordial spectrum
of GWs by making the large frequency part blue tilted. More concretely,
the GW spectrum today can be written as \citep{Figueroa:2019paj,Haque:2021dha}
 
\begin{eqnarray}
\Omega_{\gw}\left(f\right) & \simeq & \Omega_{\mathrm{\gw}}^{\rd}\times\begin{cases}
1 & f<f_{\reh}\\
\mathcal{A}_{\mathrm{s}}\left(\frac{f}{f_{\reh}}\right)^{-2\frac{1-3w_{\phi}}{1+3w_{\phi}}} & f>f_{\reh}
\end{cases}\,,\label{Ogw}
\end{eqnarray}
where $f$ denotes the present day frequency of GW modes in units
of Hz and $f_{\reh}$ is the frequency of the mode that reenters the
horizon at the end of reheating. $\Omega_{\mathrm{\gw}}^{\rd}$ is
the present day amplitude of GW modes that reenter the horizon after
reheating, that is, during the radiation domination. Because the radiation
dominated background scales with the same power as GWs at this epoch,
the slope of the GW spectrum is not modified for such frequencies.
Hence, $\Omega_{\mathrm{\gw}}^{\rd}$ is given by 
\begin{eqnarray}
\Omega_{\mathrm{\gw}}^{\rd} & \simeq & \mathcal{G}_{k}\frac{\Omega_{\r}}{36\pi^{2}}\frac{\rho_{\inf}}{\mpl^{4}}\,.
\end{eqnarray}
In Ref.~\citep{Figueroa:2019paj} $\mathcal{G}_{k}$ is approximated
by $\mathcal{G}_{k}\simeq0.39$, which we also adopt in this work.
The $w_{\phi}$ dependent constant $\mathcal{A}_{\mathrm{s}}$ is
given by
\begin{eqnarray}
\mathcal{A}_{\mathrm{s}} & = & \frac{1}{\pi}\left(1+3w_{\phi}\right)^{\frac{4}{1+3w_{\phi}}}\Gamma^{2}\left(\frac{5+3w_{\phi}}{2\left(1+3w_{\phi}\right)}\right)\,,
\end{eqnarray}
where $\Gamma$ denotes the Gamma function. In eq.~(\ref{Ogw}) we
do not include the modes that reenter the horizon after matter-radiation
equality. These modes are of a very low frequency and will play no
role in our analysis.

As can be seen from eq.~(\ref{xreh}), the DM abundance today fixes
the ratio between the reheating temperature and DM particle mass.
In its turn, knowing $T_{\reh}$, we can compute $k_{\reh}\equiv a_{\reh}H_{\reh}$
and therefore the corresponding frequency $f_{\reh}$ of that mode
today \citep{Figueroa:2019paj}  
\begin{eqnarray}
T_{\reh} & = & 9\times10^{-12}\left(\frac{200}{g_{*}}\right)^{\frac{1}{4}}\left(\frac{f_{\reh}}{\text{Hz}}\right)\mpl\,,\label{f2T}
\end{eqnarray}
where we used $\mathcal{G}_{k}\simeq0.39$ as stated above. We can
invert this expression and write  
\begin{eqnarray}
f_{\reh} & = & \frac{10^{11}}{x_{\reh}}\left(\frac{g_{*}}{200}\right)^{\frac{1}{4}}\frac{M_{\dm}}{\mpl}\,\text{ Hz}\,.\label{M2f}
\end{eqnarray}
where $x_{\reh}$ is given in eq.~(\ref{xreh}). The above expression
is another main result of this work, which relates the DM mass with
the GW frequency corresponding to reheating.

\subsection{The Bound from the Big Bang Nucleosynthesis}

The epoch of the Big Bang Nucleosynthesis (BBN) can be used to put
some of the tightest bounds on new physics at that period. Any modification
of the standard scenario results in different ratios of light element
abundances in the Universe, which are tightly constrained by observations
\citep{ParticleDataGroup:2022pth}. In particular, gravitons are relativistic
species. If their contribution to the overall energy density during
BBN is too large, the thermal history of the Universe changes and
the very successful standard BBN predictions are modified. This limits
the amount of GWs at BBN. The bound becomes especially severe for
strongly blue tilted spectra \citep{Ferreira:1997hj,Tashiro:2003qp}.

To compute the total energy density of GWs we can integrate the spectral
energy distribution over all frequencies
\begin{eqnarray}
\Omega_{\gw}^{\bbn} & \equiv & \intop_{f_{\bbn}}^{f_{\max}}\Omega_{\gw}\left(f\right)\frac{\mathrm{d}f}{f}\,,\label{Ointg}
\end{eqnarray}
where $\Omega_{\gw}\left(f\right)$ is the GW density parameter today.
$f_{\bbn}$ corresponds to the frequency of modes that reenter the
horizon at BBN. The largest frequency $f_{\max}$ is taken for $k_{\max}$
modes that exit the horizon at the end of inflation. To leave BBN
predictions intact, the upper limit  $\Omega_{\gw}^{\bbn}h^{2}<1.12\times10^{-6}$
must be satisfied \citep{Caprini:2018mtu}. The function $\Omega_{\gw}\left(f\right)$
is provided in eq.~(\ref{Ogw}). As can be seen, in the region $f>f_{\reh}$
the spectrum is blue tilted for $w_{\phi}>1/3$. Thus, in practice,
the integral in eq.~(\ref{Ointg}) is dominated by the largest frequency
modes. Using this fact, we can write the bound in eq.~(\ref{Ointg})
as \citep{Figueroa:2019paj}
\begin{eqnarray}
\Omega_{\mathrm{GW}}\left(f_{\max}\right)h^{2} & < & \Omega_{\gw}^{\bbn}h^{2}=2.24\times10^{-6}\times\frac{3w_{\phi}-1}{3w_{\phi}+1}\,.\label{BBNbound}
\end{eqnarray}

The value of $f_{\max}$ can be computed in terms of $f_{\reh}$.
By construction we have $f_{\max}/f_{\reh}=k_{\max}/k_{\reh}=H_{\inf}/a_{\reh}H_{\reh}$,
where the Hubble parameter scales as in eq.~(\ref{H}) during reheating.
Using the Friedman equation $\rho_{\inf}=3\mpl^{2}H_{\inf}^{2}$ and
eq.~(\ref{rhoT}) we find 

\begin{eqnarray}
\frac{f_{\max}}{f_{\reh}} & = & \left(4\times10^{10}\:\frac{\rho_{\inf}^{1/4}}{\mpl}\frac{\text{Hz}}{f_{\reh}}\right)^{\frac{2\left(1+3w_{\phi}\right)}{3\left(1+w_{\phi}\right)}}\,,
\end{eqnarray}
where we also employed eq.~(\ref{f2T}). Plugging this value into
the bound in eq.~(\ref{BBNbound}), we finally obtain the upper limit
of the inflation energy scale that is consistent with BBN bound
\begin{eqnarray}
\frac{\rho_{\inf}^{1/4}}{\mpl} & < & \left(\frac{36\pi^{2}}{\mathcal{G}_{k}\mathcal{A}_{\mathrm{s}}}\frac{\Omega_{\gw}^{\bbn}}{\Omega_{\r}}\right)^{\frac{3\left(1+w_{\phi}\right)}{8\left(1+3w_{\phi}\right)}}\left(2\times10^{-11}\frac{f_{\reh}}{\text{Hz}}\right)^{\frac{3w_{\phi}-1}{2\left(1+3w_{\phi}\right)}}\,.\label{V-BBN-bound}
\end{eqnarray}

\section{The Parameter Space\label{sec:DMGW}}

To find models that are compatible with observations we scan over
the range of $M_{\dm}$ and $\rho_{\inf}$ parameters for given values
of $w_{\phi}$ and $\nu$. We also identify the regions of this parameter
space that could be probed by the future GW observatories. Bellow
we explain how the upper and lower limits of $M_{\dm}$ and $\rho_{\inf}$
are computed. 

First, to compute eq.~(\ref{frin-last}) we assumed the lower bound
of $M_{\dm}$ to be $M_{\dm}>T_{\reh}$. But to ensure that the universe
is radiation dominated at BBN we must also take $T_{\reh}>T_{\bbn}\simeq1\text{ MeV}$.
Conveniently, the bound $M_{\dm}>T_{\reh}>T_{\bbn}$ can be written
as
\begin{equation}
\begin{aligned}x_{\reh} & >1\,;\\
M_{\dm} & >x_{\mathrm{\reh}}T_{\bbn}\,.
\end{aligned}
\label{MTBBN}
\end{equation}
The second inequality provides us with the lower bound for $M_{\dm}$
values. As we will see bellow, in some cases this allows for very
small $M_{\dm}$, which, one might worry, must be excluded by DM detection
experiments. However, that depends on the DM cross section. To remain
generic, we do not take into account such experiments in our analysis.

To set the upper bound, we note that to compute eq.~(\ref{frin-last}),
we also assumed $M_{\dm}<T_{\max}$. For larger $M_{\dm}$ values
DM production is exponentially suppressed. Using the value of $x_{\max}$
in eq.~(\ref{xmax-xreh}) we can write this bound as 
\begin{eqnarray}
M_{\dm} & < & x_{\reh}\left(\frac{30\rho_{\inf}}{\pi^{2}g_{*}}\right)^{\frac{1}{4}}\left(\frac{\left|\alpha\right|^{\frac{1}{4}}}{x_{\reh}a_{\max}^{1-\alpha}}\right)^{\frac{3\left(1+w_{\s}\right)}{4\left(1-\vartheta\right)}}\,.\label{MDM-max-rho}
\end{eqnarray}

As can be seen from the above inequality, the largest allowed DM particle
mass depends on the energy scale of inflation $\rho_{\inf}$. There
are two possible upper limits for $\rho_{\inf}$. First, $\rho_{\inf}$
is limited by the observational constraints on the \chnga{GWs} at
the pivot scale $k_{*}=0.05\text{ Mpc}^{-1}$, that are measured from
CMB observations \citep{BICEP:2021xfz}. This constraint is usually
expressed in terms of the tensor-to-scalar ratio 
\begin{eqnarray}
r & \equiv & \frac{A_{h}}{A_{\mathrm{s}}}\,,\label{r}
\end{eqnarray}
where $A_{h}$ and $A_{\mathrm{s}}$ are the amplitudes of the tensor
and scalar spectra at $k_{*}$ respectively. $A_{h}$ is defined as
$A_{h}=\mathcal{P}_{h}\left(k_{*}\right)$, where $\mathcal{P}_{h}$
is given in eq.~(\ref{Ph}) and $A_{\mathrm{s}}$ is determined by
the Planck normalisation $\ln\left(10^{10}A_{\mathrm{s}}\right)=3.043\pm0.014$
\citep{Planck:2018vyg}. Using eq.~(\ref{Ph}) we can write the inflation
energy scale $\rho_{*}$ as
\begin{eqnarray}
\rho_{*} & = & \frac{3}{2}r\pi^{2}A_{\mathrm{s}}\mpl^{4}\,.
\end{eqnarray}
At the end of inflation $\rho_{\inf}<\rho_{*}$. Hence, plugging in
$r<0.036$ \citep{BICEP:2021xfz} results in 
\begin{eqnarray}
\rho_{\inf}^{1/4} & < & 5.8\times10^{-3}\mpl\,.\label{rhoIup}
\end{eqnarray}

The above constraint might not be the tightest one. In certain parameter
regions, the requirement that GW do not spoil the predictions of BBN,
as it is expressed in eq.~(\ref{V-BBN-bound}), might provide more
stringent constraint. Hence, the maximum value of $\rho_{\inf}$ will
be determined by the tighter bound of the two.

Plugging the upper bound of $\rho_{\inf}$ into eq.~(\ref{MDM-max-rho}),
we can find the maximum value of allowed DM particle mass. If the
constraint in eq.~(\ref{rhoIup}) is tighter than the one in eq.~(\ref{V-BBN-bound}),
the upper bound on $M_{\dm}$ is 
\begin{eqnarray}
\frac{M_{\dm}}{\mpl} & < & 5.8\times10^{-3}x_{\reh}\left(\frac{30}{\pi^{2}g_{*}}\right)^{\frac{1}{4}}\left(\frac{\left|\alpha\right|^{\frac{1}{4}}}{x_{\reh}a_{\max}^{1-\alpha}}\right)^{\frac{3\left(1+w_{\s}\right)}{4\left(1-\vartheta\right)}}\,.\label{Mup1}
\end{eqnarray}
In the opposite case, plugging the upper value of eq.~(\ref{V-BBN-bound})
into (\ref{MDM-max-rho}), we find 
\begin{eqnarray}
\frac{M_{\dm}}{\mpl} & < & 0.81^{-\frac{5+9w_{\phi}}{3\left(1+w_{\phi}\right)}}x_{\reh}\left(\frac{1080}{g_{*}\mathcal{G}_{k}\mathcal{A}_{\mathrm{s}}}\frac{\Omega_{\gw}^{\bbn}}{\Omega_{\r}}\right)^{\frac{1}{4}}\left(\frac{\left|\alpha\right|^{\frac{1}{4}}}{x_{\reh}a_{\max}^{1-\alpha}}\right)^{\frac{2\left(1+3w_{\phi}\right)}{4\left(1-\vartheta\right)}}\,.\label{Mup2}
\end{eqnarray}

In the range of $M_{\dm}$ values that satisfy the lower limit in
eq.~(\ref{MTBBN}) and the upper limits in eq.~(\ref{Mup1}) and
(\ref{Mup2}), the lower value of $\rho_{\inf}$ can be computed by
inverting eq. (\ref{MDM-max-rho}). That is
\begin{eqnarray}
\rho_{\inf}^{\frac{1}{4}} & > & \frac{M_{\dm}}{x_{\reh}}\left(\frac{\pi^{2}g_{*}}{30}\right)^{\frac{1}{4}}\left(\frac{x_{\reh}a_{\max}^{1-\alpha}}{\left|\alpha\right|^{\frac{1}{4}}}\right)^{\frac{3\left(1+w_{\s}\right)}{4\left(1-\vartheta\right)}}\label{rho-low}
\end{eqnarray}

\begin{figure}
\begin{centering}
\includegraphics[scale=0.6]{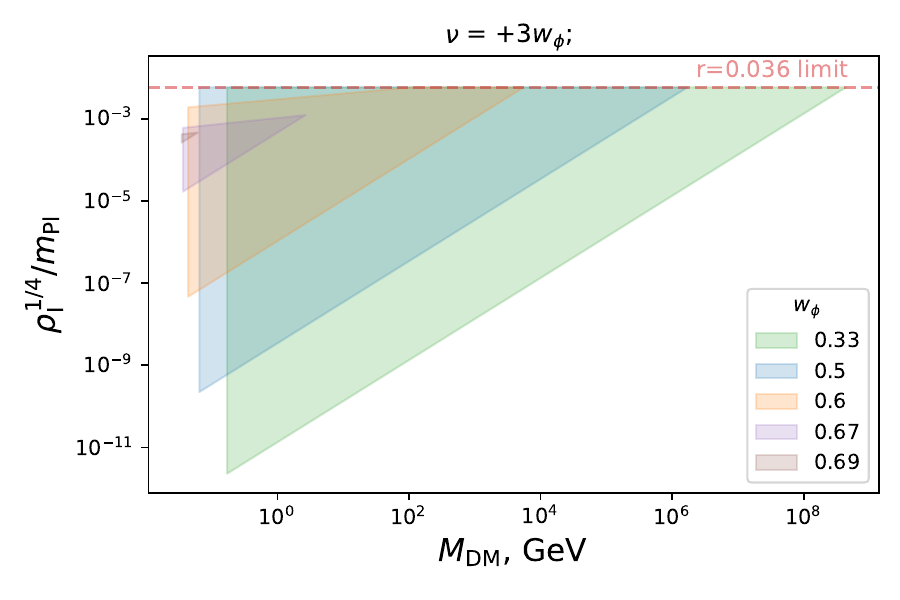}
\par\end{centering}
\caption{\label{fig:mrho}The parameter range with allowed values of $M_{\protect\dm}$
and $\rho_{\protect\inf}$. The dashed blue horizontal line represents
the upper limit on $\rho_{\protect\inf}$ given in eq.~(\ref{rhoIup}).
The actual upper limit of $\rho_{I}$ is constrained either by this
bound or the BBN bound in eq.~(\ref{V-BBN-bound}). The lower limit
of $\rho_{\protect\inf}$ is determined by the condition in eq.~(\ref{rho-low}).
The lower limit of $M_{\protect\dm}$ is given in eq.~\ref{MTBBN}
and the upper one in eq.~(\ref{Mup1}) or (\ref{Mup2}), whichever
is the tighter one.}
\end{figure}

To explore the parameter space of this model we adopt the following
values
\begin{eqnarray}
g_{*} & = & 200\,,\label{gp}\\
M_{\dm}^{2}\left\langle \sigma\left|v\right|\right\rangle  & = & 10^{-2}\,.\label{Mp}
\end{eqnarray}
As for the values of $\nu$ we consider two possibilities, as mentioned
in subsection \ref{subsec:Setup}. If the dominant inflaton decay
branching ratio is determined by the $\phi\rightarrow\psi\bar{\psi}$
process, where $\psi$ is a spin-$\frac{1}{2}$ particles, then $\nu=-3w_{\phi}$
\citep{Shtanov:1994ce}. However, in this case we do not find a parameter
region which would result in the right abundance of DM. \todo[inline]{There are various ways to show this, but one can see that}.
In the other case, when the dominant decay channel is $\phi\rightarrow\chi^{2}$
and/or the scattering $\phi^{2}\rightarrow\chi^{2}$, the decay rate
scaling is given by $\nu=+3w_{\phi}$. This scenario can be successfully
realised in a limited range of $w_{\phi}$ values, which is demonstrated
in Figure~\ref{fig:mrho}. 

We are also interested in the possibility to probe such models with
GW observations. To this goal we compute GW spectra for all the allowed
range of $M_{\dm}$, $\rho_{\inf}$ and $w_{\phi}$ values and inspect
if they cross the sensitivity curves of several GW observatories such
as aLIGO, aVirgo, the Kamioka Gravitational-Wave Detector (KAGRA)
\citep{Somiya:2011np,Aso:2013eba,KAGRA:2018plz,KAGRA:2019htd}, NANOGrav
\citep{McLaughlin:2013ira,NANOGRAV:2018hou,Brazier:2019mmu}, PPTA
\citep{Manchester:2012za,Shannon:2015ect}, IPTA \citep{Hobbs:2009yy,Manchester:2013ndt,Verbiest:2016vem,Hazboun:2018wpv},
SKA \citep{Carilli:2004nx,Janssen:2014dka,Weltman:2018zrl}, LISA
\citep{LISA:2017pwj,Baker:2019nia}, BBO \citep{Crowder:2005nr,Corbin:2005ny,Harry:2006fi},
DECIGO \citep{Seto:2001qf,Kawamura:2006up,Yagi:2011wg}, CE \citep{LIGOScientific:2016wof,Reitze:2019iox},
ET \citep{Punturo:2010zz,Hild:2010id,Sathyaprakash:2012jk,Maggiore:2019uih}.
The power-law-integrated sensitivity curves \citep{Thrane:2013oya}
of these observatories are taken from \citep{Schmitz:2020syl}.

\begin{figure}
\includegraphics[scale=0.6]{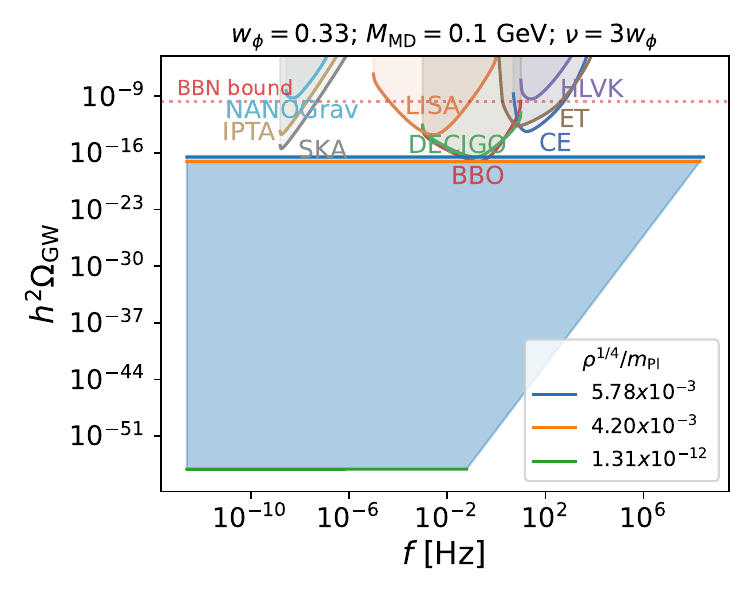}~ \includegraphics[scale=0.6]{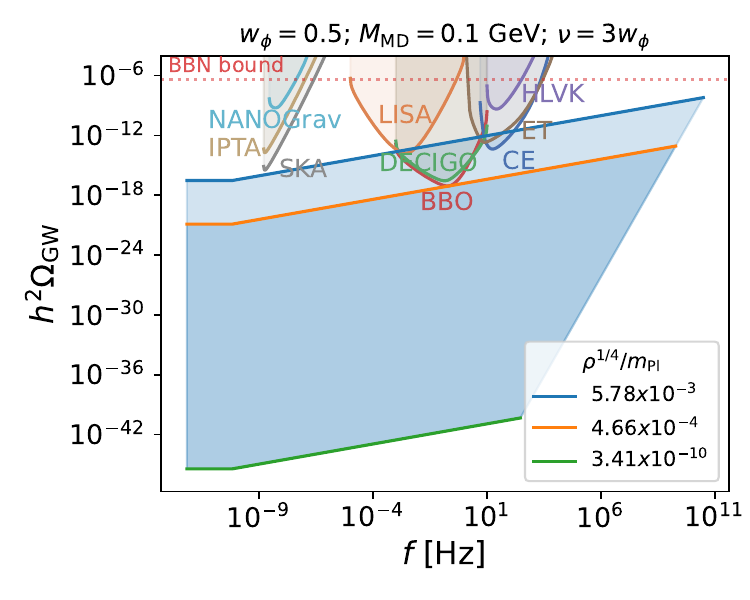}

\includegraphics[scale=0.6]{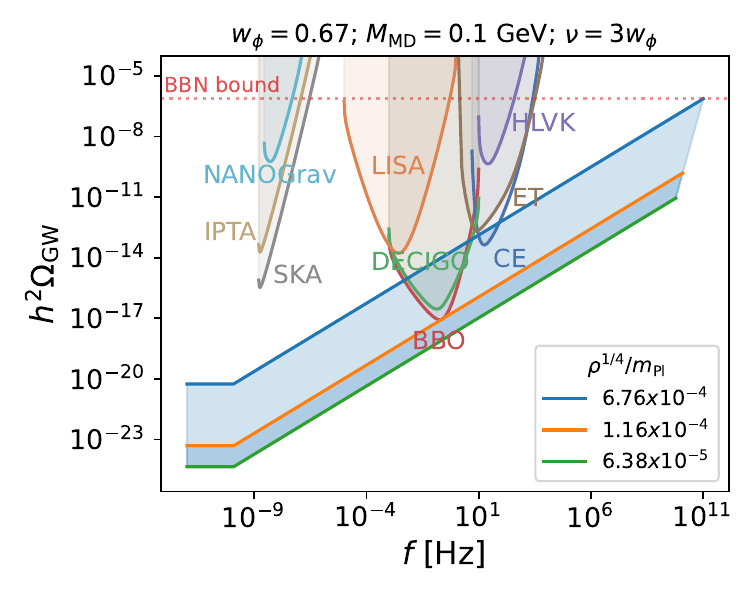}

\caption{\label{fig:spectra}A few examples of GWs spectra for several freeze-in
models of DM with three values of $w_{\phi}$ and three DM particle
masses. Each spectrum corresponds to one curve. The lowest (green)
curve denotes a model with the minimal inflation energy scale that
is allowed by eq.~(\ref{rho-low}). The upper (blue) curve corresponds
to the spectrum of the maximal allowed inflationary scale. In the
first two subplots this corresponds to the bound in eq.~(\ref{rhoIup}).
In the last subplot the maximal value of $\rho_{\protect\inf}$ is
given by the inequality in eq.~(\ref{V-BBN-bound}). The shaded region
denotes the range of spectra that are allowed by the aforementioned
bounds, the lighter part denoting the potentially observable range.
In all of these models we take $g_{*}=200$, $M_{\protect\dm}\left\langle \sigma\left|v\right|\right\rangle =10^{-2}$.
We also show the sensitivity curves of the present and future planned
GW detectors.}
\end{figure}

We demonstrate GW spectra in Figure~\ref{fig:spectra}. In this figure
the blue shaded regions denote the range of spectra $\Omega_{\gw}\left(f\right)h^{2}$
for a fixed value of $w_{\phi}$ and $M_{\dm}$, while $\rho_{\inf}$
varies within the above discussed limits. In these plots one can also
see sensitivity curves of GW observatories. The lighter shaded part
of the blue region denotes the range of inflation energy scale $\rho_{\inf}$
that results in observable GW. In the first plot of Figure~\ref{fig:spectra}
we show the spectrum for $n=2$. As expected, it is flat. The third
plot serves to illustrate the parameter values for which the maximum
value of $\rho_{\inf}$ is limited not by eq.~(\ref{rhoIup}) but
by the BBN bound in eq.~(\ref{V-BBN-bound}).

\begin{figure}
\begin{centering}
\includegraphics[scale=0.55]{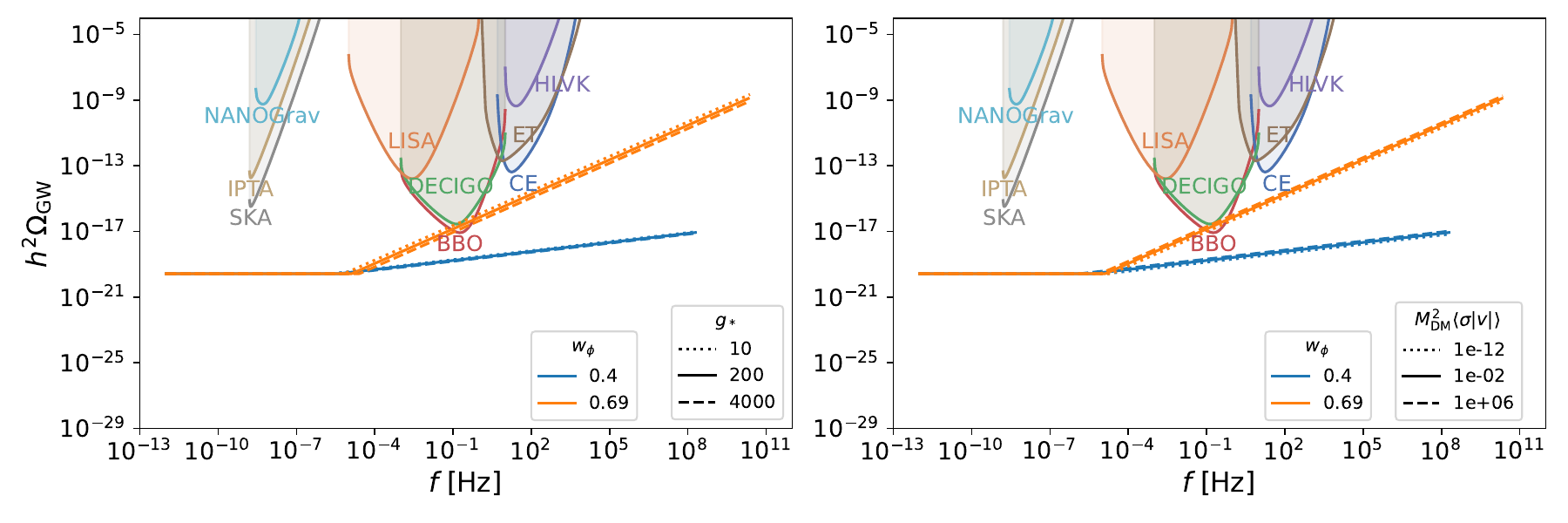}
\par\end{centering}
\caption{\label{fig:sens}The sensitivity of the GW spectrum to the effective
number of relativistic degrees of freedom during reheating, $g_{*}$,
and the thermal average cross section for annihilation of DM, $\left\langle \sigma\left|v\right|\right\rangle $.
For all models we use $M_{\protect\dm}=10^{4}\text{ GeV}$ and $\rho_{\protect\inf}^{1/4}=10^{-3}\protect\mpl$.}
\end{figure}

The spectra in Figure~\ref{fig:spectra} are calculated for fixed
values of $g_{*}$ and $M_{\dm}^{2}\left\langle \sigma\left|v\right|\right\rangle $
given in eqs.~(\ref{gp}) and (\ref{Mp}). However, the results are
very insensitive to the exact values of these parameters, as they
are suppressed by the power $2\left(1-\vartheta\right)/\left(10\vartheta-1+3w_{\phi}\right)$
in eq.~(\ref{xreh}). To demonstrate this fact, we provide Figure~\ref{fig:sens},
where several spectra are shown with various values of $w_{\phi}$,
$g_{*}$ and $M_{\dm}^{2}\left\langle \sigma\left|v\right|\right\rangle $
(and $\nu=3w_{\phi}$).

\begin{figure}
\begin{centering}
\includegraphics[scale=0.6]{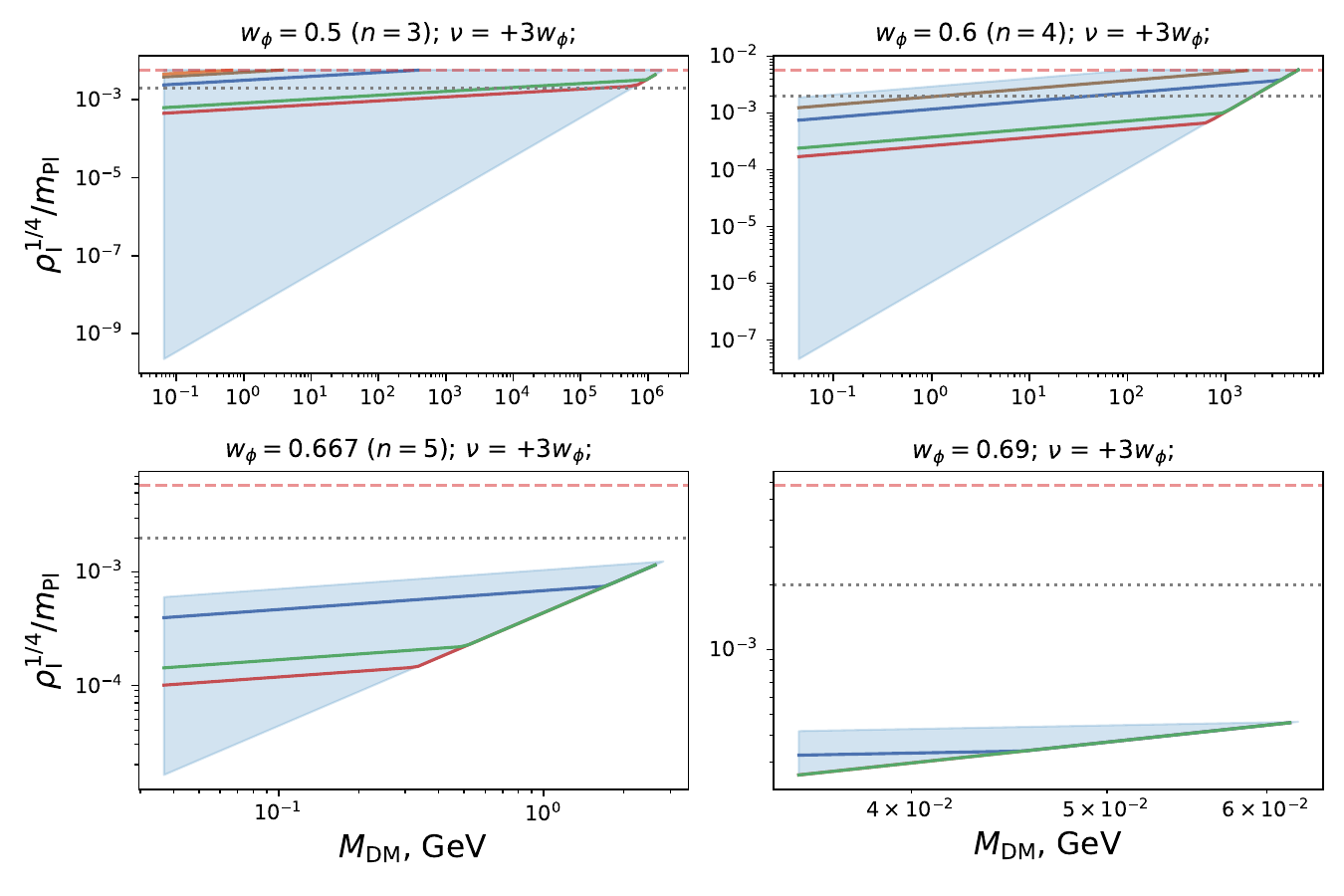}
\par\end{centering}
\caption{\label{fig:prmsp}Parameter ranges (blue regions) for $\rho_{\protect\inf}$
and $M_{\protect\dm}$ for various values of $w_{\phi}$ during reheating.
The red dashed line corresponds to the $\rho_{\protect\inf}$ bound
in eq.~(\ref{rhoIup}) and the grey dotted line corresponds to the
sensitivity limit of CMB Stage-4 project. The colour coded curves
mark the regions that will be detectable by future GW observatories.
Such regions are above the curves, where the colours are the same
as of sensitivity curves in Figure \ref{fig:spectra}. Notice, that
there is no allowed parameter range for $n=6$ or large. The lower
right plot is just for illustrative purposes, it does not correspond
to any integer value of $n$.}
\end{figure}

A full scan of the $M_{\dm}$ and $\rho_{\inf}$ values is shown in
Figure~\ref{fig:prmsp}. In this figure $M_{\dm}$ varies from the
value given in eq.~(\ref{MTBBN}) to the value given in either eq.~(\ref{Mup1})
or eq.~(\ref{Mup2}), whichever is smaller. The lower limit of $\rho_{\inf}$
is determined by eq.~(\ref{rho-low}) and the upper limit by the
tighter constraint of the one given in eq.~(\ref{V-BBN-bound}) or
(\ref{rhoIup}). We perform this scan for several values of $w_{\phi}=0.5,\:0.6,\:0.667$
and $0.69$. The first three correspond to $n=3,\:4,$ and $5$ in
eq.~(\ref{V}) respectively. Models with any larger value of $n$
are excluded. $w_{\phi}=0.69$ is plotted just for illustrative purposes,
it does not correspond to any integer $n$. In this plot we also display
regions that will be accessible by future GW observatories. They are
denoted by coloured curves. Regions above a given curve falls within
the sensitivity curve of the experiment. The colour coding is the
same as in Figure~\ref{fig:spectra}. We also display the predicted
sensitivity of the CMB Stage-4 project \citep{Abazajian:2019eic}.
As can be seen from the figure, within some parameter range the detection
of GWs can establish constraints on the inflation energy scale that
are tighter by several orders of magnitude, depending on the value
of $M_{\dm}$. This is true at least for slow-roll inflation, where
the energy scale of inflation does not change much. This is due to
the fact that $\rho_{\inf}$ in Figure~\ref{fig:prmsp} corresponds
to the energy density at the end of inflation, while CMB polarisation
measurements, such as CMB Stage-4, constrain the epoch that corresponds
to tens of e-folds before that.

\section{Summary and Conclusions}

The properties of DM particles, that are required to reproduce the
present day abundance, depend on the thermal history of the Universe.
This is especially true if DM is produced via the so called freeze-in
mechanism. We can use this sensitivity to constrain the earliest epochs
of the evolution of the Universe. 

In this work we study the freeze-in DM production mechanism and how
it is affected by the non-standard expansion history during reheating.
In order to reproduce the observed present day abundance, DM particle
mass is related to the reheating temperature as shown in eq.~(\ref{xreh}).
The relation, among other parameters, also depends on the effective
equation of state, $w_{\phi}$, during reheating. To fix that value
another set of observations is needed, which is provided by GW observatories.

Observations of GWs give us a new tool to study the Universe and the
very early stages of its evolution. Quantum vacuum fluctuations during
inflation leave the Universe filled with a stochastic GW background.
The primordial spectrum of this background is (quasi-)flat. The ensuing
evolution modifies a part of that spectrum. Any period where the expansion
rate of the Universe deviates from the radiation dominated one, tilts
the spectrum within the range of subhorizon modes. In particular,
the expansion rate that is faster than the radiation dominated one,
makes the corresponding part of the spectrum blue tilted. Depending
on the magnitude of the deviation and the scale of inflation, the
tilt can be sufficiently large to make such GW signal directly observable.
In such cases, we can use eq.~(\ref{xreh}) to constrain the inflation
energy scale, the reheating temperature and the expansion rate during
the reheating.

In Figure~\ref{fig:spectra} we show how the spectrum of GWs is related
to the main three parameters of the model: DM particle mass $M_{\dm}$,
the expansion rate during reheating and inflation energy scale. In
principle, the spectrum depends on the number of relativistic degrees
of freedom during reheating, $g_{*}$, and the thermal average cross
section for annihilation of DM $\left\langle \sigma\left|v\right|\right\rangle $.
However, as we show in Figure~\ref{fig:sens}, this dependence is
very strongly suppressed.

Finally, we show the allowed parameter space of such models for several
expansion rates (parametrised by $w_{\phi}$) in Figure~\ref{fig:prmsp}.
We can see that in the case of large enough inflation energy scale
and $3\le n\le5$ $\left(1/2\le w_{\phi}\le2/3\right)$, such models
produce GWs that are detectable by future observatories. We can also
look at this result differently: if DM are produced by the freeze-in
process, the measurement of the $M_{\dm}$ and the GW spectrum will
allow us to constrain the reheating temperature and inflation energy
scale. For some values of $M_{\dm}$ the latter constraint can more
than an order of magnitude tighter than the predicted bound from the
future CMB Stage-4 project.
\begin{acknowledgments}
M.K. is partially supported by the María Zambrano grant, provided
by the Ministry of Universities from the Next Generation funds of
the European Union. This work is also partially supported by the MICINN
(Spain) projects PID2019-107394GB-I00/AEI/10.13039/501100011033 and
PID2022-139841NB-I00 (AEI/FEDER, UE), COST (European Cooperation in
Science and Technology) Actions CA21106 and CA21136.
\end{acknowledgments}

\bibliographystyle{aipnum4-2}
\bibliography{freezein-gw.bbl}

\end{document}